# Analysis of Sorting Algorithms by Kolmogorov Complexity (A Survey)

Paul Vitányi [*]

November 5, 2018

**Abstract**

Recently, many results on the computational complexity of sorting algorithms were obtained using Kolmogorov complexity (the incompressibility method). Especially, the usually hard average-case analysis is ammenable to this method. Here we survey such results about Bubblesort, Heapsort, Shellsort, Dobosiewicz-sort, Shakersort, and sorting with stacks and queues in sequential or parallel mode. Especially in the case of Shellsort the uses of Kolmogorov complexity surprisingly easily resolved problems that had stayed open for a long time despite strenuous attacks.

## 1 Introduction

We survey recent results in the analysis of sorting algorithms using a new technical tool: the incompressibility method based on Kolmogorov complexity. Complementing approaches such as the counting method and the probabilistic method, the new method is especially suited for the average-case analysis of algorithms and machine models, whereas average-case analysis is usually more difficult than worst-case analysis using the traditional methods. Obviously, the results described can be obtained using other proof methods—all true provable statements must be provable from the axioms of mathematics by the inference methods of mathematics. The question is whether a particular proof method facilitates and guides the proving effort. The following examples make clear that thinking in terms of coding and the incompressibility method suggests simple proofs that resolve long-standing open problems. A survey of the use of the incompressibility method in combinatorics, computational complexity, and the analysis of algorithms is [16] Chapter 6, and other recent work is [2, 15].

We give some definitions to establish notation. For introduction, details, and proofs, see [16]. We write *string* to mean a finite binary string. Other finite objects can be encoded into strings in natural ways. The set of strings is denoted by $\{0,1\}^*$. Let $x, y, z \in \mathcal{N}$, where $\mathcal{N}$ denotes the set of natural numbers. Identify $\mathcal{N}$ and $\{0,1\}^*$ according to the correspondence

$$(0, \epsilon), (1, 0), (2, 1), (3, 00), (4, 01), \ldots.$$

Here $\epsilon$ denotes the *empty word* with no letters. The *length* of $x$ is the number of bits in the binary string $x$ and is denoted by $l(x)$. For example, $l(010) = 3$ and $l(\epsilon) = 0$. The emphasis is on binary sequences only for convenience; observations in every alphabet can be so encoded in a way that is 'theory neutral'.

---

[*]CWI, Kruislaan 413, 1098 SJ Amsterdam, The Netherlands. Email: paulv@cwi.nl. Supported in part via NeuroCOLT II ESPRIT Working Group.

**Self-delimiting Codes:** A binary string $y$ is a *proper prefix* of a binary string $x$ if we can write $x = yz$ for $z \neq \epsilon$. A set $\{x, y, \ldots\} \subseteq \{0,1\}^*$ is *prefix-free* if for every pair of distinct elements in the set neither is a proper prefix of the other. A prefix-free set is also called a *prefix code*. Each binary string $x = x_1 x_2 \ldots x_n$ has a special type of prefix code, called a *self-delimiting code*,

$$\bar{x} = 1^n 0 x_1 x_2 \ldots x_n.$$

This code is self-delimiting because we can effectively determine where the code word $\bar{x}$ ends by reading it from left to right without backing up. Using this code we define the standard self-delimiting code for $x$ to be $x' = \overline{l(x)}x$. It is easy to check that $l(\bar{x}) = 2n+1$ and $l(x') = n + 2\log n + 1$.

Let $\langle \cdot, \cdot \rangle$ be a standard one-one mapping from $\mathcal{N} \times \mathcal{N}$ to $\mathcal{N}$, for technical reasons chosen such that $l(\langle x, y \rangle) = l(y) + l(x) + 2l(l(x)) + 1$, for example $\langle x, y \rangle = x'y = 1^{l(l(x))} 0 l(x) xy$.

**Kolmogorov Complexity:** Informally, the Kolmogorov complexity, or algorithmic entropy, $C(x)$ of a string $x$ is the length (number of bits) of a shortest binary program (string) to compute $x$ on a fixed reference universal computer (such as a particular universal Turing machine). Intuitively, $C(x)$ represents the minimal amount of information required to generate $x$ by any effective process, [10]. The conditional Kolmogorov complexity $C(x \mid y)$ of $x$ relative to $y$ is defined similarly as the length of a shortest program to compute $x$, if $y$ is furnished as an auxiliary input to the computation. The functions $C(\cdot)$ and $C(\cdot \mid \cdot)$, though defined in terms of a particular machine model, are machine-independent up to an additive constant (depending on the particular enumeration of Turing machines and the particular reference universal Turing machine selected). They acquire an asymptotically universal and absolute character through Church's thesis, and from the ability of universal machines to simulate one another and execute any effective process, see for example [16]. Formally:

DEFINITION 1 Let $T_0, T_1, \ldots$ be a standard enumeration of all Turing machines. Choose a universal Turing machine $U$ that expresses its universality in the following manner:

$$U(\langle \langle i, p \rangle, y \rangle) = T_i(\langle p, y \rangle)$$

for all $i$ and $\langle p, y \rangle$, where $p$ denotes a Turing program for $T_i$ and $y$ an input. We fix $U$ as our *reference universal computer* and define the *conditional Kolmogorov complexity* of $x$ given $y$ by

$$C(x \mid y) = \min_{q \in \{0,1\}^*} \{l(q) : U(\langle q, y \rangle) = x\},$$

for every $q$ (for example $q = \langle i, p \rangle$ above) and auxiliary input $y$. The *unconditional Kolmogorov complexity* of $x$ is defined by $C(x) = C(x \mid \epsilon)$. For convenience we write $C(x, y)$ for $C(\langle x, y \rangle)$, and $C(x \mid y, z)$ for $C(x \mid \langle y, z \rangle)$.

**Incompressibility:** First we show that the Kolmogorov complexity of a string cannot be significantly more than its length. Since there is a Turing machine, say $T_i$, that computes the identity function $T_i(x) \equiv x$, and by definition of universality of $U$ we have $U(\langle i, p \rangle) = T_i(p)$. Hence, $C(x) \leq l(x) + c$ for fixed $c \leq 2\log i + 1$ and all $x$. [1] [2]

---

[1] "$2 \log i$" and not "$\log i$" since we need to encode $i$ in such a way that $U$ can determine the end of the encoding. One way to do that is to use the code $1^l(l(i))0l(i)i$ which has length $2l(l(i)) + l(i) + 1 < 2 \log i$ bits.

[2] In what follows, "log" denotes the binary logarithm. "$\lfloor r \rfloor$" is the greatest integer $q$ such that $q \leq r$.

It is easy to see that there are also strings that can be described by programs much shorter than themselves. For instance, the function defined by $f(1) = 2$ and $f(i) = 2^{f(i-1)}$ for $i > 1$ grows very fast, $f(k)$ is a "stack" of $k$ twos. It is clear that for every $k$ it is the case that $f(k)$ has complexity at most $C(k) + O(1)$. What about incompressibility? For every $n$ there are $2^n$ binary strings of length s$n$, but only $\sum_{i=0}^{n-1} 2^i = 2^n - 1$ descriptions in binary string format of lengths less than $n$. Therefore, there is at least one binary string $x$ of length $n$ such that $C(x) \geq n$. We call such strings *incompressible*. The same argument holds for conditional complexity: since for every length $n$ there are at most $2^n - 1$ binary programs of lengths $< n$, for every binary string $y$ there is a binary string $x$ of length $n$ such that $C(x \mid y) \geq n$. Strings that are incompressible are patternless, since a pattern could be used to reduce the description length. Intuitively, we think of such patternless sequences as being random, and we use "random sequence" synonymously with "incompressible sequence." There is also a formal justification for this equivalence, which does not need to concern us here. Since there are few short programs, there can be only few objects of low complexity: the number of strings of length $n$ that are compressible by at most $\delta$ bits is at least $2^n - 2^{n-\delta} + 1$.

LEMMA 1 *Let $\delta$ be a positive integer. For every fixed $y$, every set $S$ of cardinality $m$ has at least $m(1 - 2^{-\delta}) + 1$ elements $x$ with $C(x \mid y) \geq \lfloor \log m \rfloor - \delta$.*

PROOF. There are $N = \sum_{i=0}^{n-1} 2^i = 2^n - 1$ binary strings of length less than $n$. A fortiori there are at most $N$ elements of $S$ that can be computed by binary programs of length less than $n$, given $y$. This implies that at least $m - N$ elements of $S$ cannot be computed by binary programs of length less than $n$, given $y$. Substituting $n$ by $\lfloor \log m \rfloor - \delta$ together with Definition 1 yields the lemma. □

LEMMA 2 *If $A$ is a set, then for every $y$ every element $x \in A$ has complexity $C(x|A, y) \leq \log |A| + O(1)$.*

PROOF. A string $x \in A$ can be described by first describing $A$ in $O(1)$ bits and then giving the index of $x$ in the enumeration order of $A$. □

As an example, set $S = \{x : l(x) = n\}$. Then is $|S| = 2^n$. Since $C(x) \leq n + c$ for some fixed $c$ and all $x$ in $S$, Lemma 1 demonstrates that this trivial estimate is quite sharp. If we are given $S$ as an explicit table then we can simply enumerate its elements (in, say, lexicographical order) using a fixed program not depending on $S$ or $y$. Such a fixed program can be given in $O(1)$ bits. Hence the complexity satisfies $C(x \mid S, y) \leq \log |S| + O(1)$.

**Incompressibility Method:** In a typical proof using the incompressibility method, one first chooses an incompressible object from the class under discussion. The argument invariably says that if a desired property does not hold, then in contrast with the assumption, the object can be compressed significantly. This yields the required contradiction. Since most objects are almost incompressible, the desired property usually also holds for almost all objects, and hence on average. Below, we demonstrate the utility of the incompressibility method to obtain simple and elegant proofs.

**Average-case Complexity:** For many algorithms, it is difficult to analyze the average-case complexity. Generally speaking, the difficulty comes from the fact that one has to analyze the time complexity for all inputs of a given length and then compute the average. This is a difficult task. Using the incompressibility method, we choose just one input — a representative input. Via Kolmogorov complexity, we can show that the time complexity of this input is in fact the average-case complexity of all inputs of this length. Constructing such a "representative input" is impossible, but we know it exists and this is sufficient.

In average-case analysis, the incompressibility method has an advantage over a probabilistic approach. In the latter approach, one deals with expectations or variances over some ensemble of objects. Using Kolmogorov complexity, we can reason about an incompressible individual object. Because it is incompressible it has all simple statistical properties with certainty, rather than having them hold with some (high) probability as in a probabilistic analysis. This fact greatly simplifies the resulting analysis.

## 2 Bubblesort

A simple introductory example of the application of the incompressibility method is the average-case analysis of Bubblesort. The classical approach can be found in [11]. It is well-known that Bubblesort uses $\Theta(n^2)$ comparisons/exchanges on the average. We present a very simple proof of this fact. The proof is based on the following intuitive idea: There are $n!$ different permutations. Given the sorting process (the insertion paths in the right order) one can recover the correct permutation from the sorted list. Hence one requires $n!$ pairwise different sorting processes. This gives a lower bound on the minimum of the maximal length of a process. We formulate the proof in the crisp format of incompressibility. In Bubblesort we make passes from left to right over the permutation to be sorted and always move the currently largest element right by exchanges between it and the right-adjacent element—if that one is smaller. We make at most $n-1$ passes, since after moving all but one element in the correct place the single remaining element must be also in its correct place (it takes two elements to be wrongly placed). The total number of exchanges is obviously at most $n^2$, so we only need to consider the lower bound. Let $B$ be a Bubblesort algorithm. For a permutation $\pi$ of the elements $1, \ldots, n$, we can describe the total number of exchanges by $M := \sum_{i=1}^{n-1} m_i$ where $m_i$ is the initial distance of element $n-i$ to its final position. Note that in every pass more than one element may "bubble" right but that means simply that in the future passes of the sorting process an equal number of exchanges will be saved for the element to reach its final position. That is, every element executes a number of exchanges going right that equals precisely the initial distance between its start position to its final position. It is clear that $M \leq n^2$ for all permutations. Given $m_1, \ldots, m_{n-1}$, in that order, we can reconstruct the original permutation from the final sorted list. Since choosing $a$ elements from a list of $b+a$ elements divides the remainder in a sequence of $a+1$ possibly empty sublists, there are

$$B(M) = \binom{M+n-2}{n-2}$$

possibilities to partition $M$ into $n-1$ ordered non-negative summands. Therefore, we can describe $\pi$ by $M, n$, an index of $\log B(M)$ bits to describe $m_1, \ldots, m_{n-1}$ among all partitions of $M$, and an program $P$ that reconstructs $\pi$ from these parameters and the final sorted list $1, \ldots, n$. Consider permutations $\pi$ satisfying $C(\pi \mid n, B(M), P) \geq \log n! - \log n$. Then by Lemma 2 at least a $(1 - 1/n)$th fraction of all permutations of $n$ elements have that high complexity. Under this complexity condition on $\pi$, we also have $M \geq n$. (If $M < n$ then $C(\pi \mid n, B(M), P) = O(n)$.) Since the description of $\pi$ we have constructed is effective, its length must be at least $C(\pi \mid n, B, P)$. Encoding $M$ self-delimiting, in order to be able to separate $M$ from $B(M)$ in a concatenation of the binary descriptions, we therefore find $\log M + 2 \log \log M + \log B(M) \geq n \log n - O(\log n)$. Substitute a good estimate for $\log B(M)$ (the formula used later in the Shellsort example, Section 4) divide by $n$, and discard the terms that vanish with $n$, assuming $2 < n \leq M \leq n^2$, yields $\log(1 + M/(n-2)) \geq \log n + O(1)$. By the above, this holds for at least an $(1 - 1/n)$th fraction of all permutations, and hence gives us an $\Omega(n^2)$ lower bound on the expected number of comparisons/exchanges.

## 3   Heapsort

Heapsort is a widely used sorting algorithm. One reason for its prominence is that its running time is *guaranteed* to be of order $n \log n$, and it does not require extra memory space. The method was first discovered by J.W.J. Williams, [29], and subsequently improved by R.W. Floyd [4] (see [11]). Only recently has one succeeded in giving a precise analysis of its average-case performance [23]. I. Munro has suggested a remarkably simple solution using incompressibility [18] initially reported in [16].

A "heap" can be visualized as a complete directed binary tree with possibly some rightmost nodes being removed from the deepest level. The tree has $n$ nodes, each of which is labeled with a different key, taken from a linearly ordered domain. The largest key $k_1$ is at the root (on top of the heap), and each other node is labeled with a key that is less than the key of its father.

DEFINITION 2   Let *keys* be elements of $\mathcal{N}$. An array of keys $k_1, \ldots, k_n$ is a *heap* if they are partially ordered such that
$$k_{\lfloor j/2 \rfloor} \geq k_j \text{ for } 1 \leq \lfloor j/2 \rfloor < j \leq n.$$

Thus, $k_1 \geq k_2, k_1 \geq k_3, k_2 \geq k_4$, and so on. We consider "in place" sorting of $n$ keys in an array $A[1..n]$ without use of additional memory.

**Heapsort** {Initially, $A[1..n]$ contains $n$ keys. After sorting is completed, the keys in $A$ will be ordered as $A[1] < A[2] < \cdots < A[n]$.}

**Heapify:** {Regard $A$ as a tree: the root is in $A[1]$; the two sons of $A[i]$ are at $A[2i]$ and $A[2i+1]$, when $2i, 2i+1 \leq n$. We convert the tree in $A$ to a heap.} **Repeat for** $i = \lfloor n/2 \rfloor, \lfloor n/2 \rfloor - 1, \ldots, 1$: {the subtree rooted at $A[i]$ is now almost a heap except for $A[i]$} push the key, say $k$, at $A[i]$ down the tree (determine which of the two sons of $A[i]$ possesses the greatest key, say $k'$ in son $A[2i+j]$ with $j$ equals 0 or 1); **if** $k' > k$ **then** put $k$ in $A[2i+j]$ and **repeat** this process pushing $k'$ at $A[2i+j]$ down the tree **until** the process reaches a node that does not have a son whose key is greater than the key now at the father node.

**Sort: Repeat for** $i = n, n-1, \ldots, 2$: {$A[1..i]$ contains the remaining heap and $A[i+1..n]$ contains the already sorted list $k_{i+1}, \ldots, k_n$ of largest elements. By definition, the element on top of the heap in $A[1]$ must be $k_i$.} switch the key $k_i$ in $A[1]$ with the key $k$ in $A[i]$, extending the sorted list to $A[i..n]$. Rearrange $A[1..i-1]$ to a heap with the largest element at $A[1]$.

It is well known that the Heapify step can be performed in $O(n)$ time. It is also known that the Sort step takes no more than $O(n \log n)$ time. We analyze the precise average-case complexity of the Sort step. There are two ways of rearranging the heap: Williams's method and Floyd's method.

**Williams's Method:** {Initially, $A[1] = k$.}

**Repeat** compare the keys of $k$'s two direct descendants; **if** $m$ is the larger of the two **then** compare $k$ and $m$; **if** $k < m$ **then** switch $k$ and $m$ in $A[1..i-1]$ **until** $k \geq m$.

**Floyd's Method:** {Initially, $A[1]$ is empty.} Set $j := 1$;

**while** $A[j]$ is not a leaf **do**:
     **if** $A[2j] > A[2j+1]$ **then** $j := 2j$
     **else** $j := 2j + 1$;

**while** $k > A[j]$ **do**:

    {back up the tree until the correct position for $k$} $j := \lfloor j/2 \rfloor$;

**move** keys of $A[j]$ and each of its ancestors one node upwards;

    Set $A[j] := k$.

The difference between the two methods is as follows. Williams's method goes from the root at the top down the heap. It makes two comparisons with the son nodes and one data movement at each step until the key $k$ reaches its final position. Floyd's method first goes from the root at the top down the heap to a leaf, making only one comparison each step. Subsequently, it goes from the bottom of the heap up the tree, making one comparison each step, until it finds the final position for key $k$. Then it moves the keys, shifting every ancestor of $k$ one step up the tree. The final positions in the two methods are the same; therefore both algorithms make the same number of key movements. Note that in the last step of Floyd's algorithm, one needs to move the keys carefully upward the tree, avoiding swaps that would double the number of moves.

The heap is of height $\log n$. If Williams's method uses $2d$ comparisons, then Floyd's method uses $d + 2\delta$ comparisons, where $\delta = \log n - d$. Intuitively, $\delta$ is generally very small, since most elements tend to be near the bottom of the heap. This makes it likely that Floyd's method performs better than Williams's method. We analyze whether this is the case. Assume a uniform probability distribution over the lists of $n$ keys, so that all input lists are equally likely.

Average-case analysis in the traditional manner suffers from the problem that, starting from a uniform distribution on the lists, it is difficult to compute the distribution on the resulting initial heaps, and increasingly more difficult to compute the distributions on the sequence of decreasing-size heaps after subsequent heapsort steps. The sequence of distributions seem somehow realated, but this is hard to express and exploit in the traditional approach. In contrast, using Kolmogorov complexity we express this similarity without having to be precise about the distributions.

THEOREM 1 *On average (uniform distribution), Heapsort makes $n \log n + O(n)$ data movements. Williams's method makes $2n \log n - O(n)$ comparisons on average. Floyd's method makes $n \log n + O(n)$ comparisons on average.*

PROOF. Given $n$ keys, there are $n!$ ($\approx n^n e^{-n} \sqrt{2\pi n}$ by Stirling's formula) permutations. Hence we can choose a permutation $p$ of $n$ keys such that

$$C(p|n) \geq n \log n - 2n, \qquad (1)$$

justified by Theorem 1, page 3. In fact, most permutations satisfy Equation 1.

CLAIM 1 *Let $h$ be the heap constructed by the **Heapify** step with input $p$ that satisfies Equation 1. Then*

$$C(h|n) \geq n \log n - 6n. \qquad (2)$$

PROOF. Assume the contrary, $C(h|n) < n \log n - 6n$. Then we show how to describe $p$, using $h$ and $n$, in fewer than $n \log n - 2n$ bits as follows. We will encode the **Heapify** process that constructs $h$ from $p$. At each loop, when we push $k = A[i]$ down the subtree, we record the path that key $k$ traveled: 0 indicates a left branch, 1 means a right branch, 2 means halt. In total, this requires $(n \log 3) \sum_j j/2^{j+1} \leq 2n \log 3$ bits. Given the final heap $h$ and the above description of updating paths, we can reverse the procedure of **Heapify** and reconstruct $p$. Hence, $C(p|n) < C(h|n) + 2n \log 3 + O(1) < n \log n - 2n$, which is a contradiction. (The term $6n$ above can be reduced by a more careful encoding and calculation.) □

We give a description of $h$ using the history of the $n-1$ heap rearrangements during the Sort step. We only need to record, for $i := n-1, \ldots, 2$, at the $(n-i+1)$st round of the Sort step, the final position where $A[i]$ is inserted into the heap. Both algorithms insert $A[i]$ into the same slot using the same number of data moves, but a different number of comparisons.

We encode such a final position by describing the path from the root to the position. A path can be represented by a sequence $s$ of 0's and 1's, with 0 indicating a left branch and 1 indicating a right branch. Each path $i$ is encoded in self-delimiting form by giving the value $\delta_i = \log n - l(s_i)$ encoded in self-delimiting binary form, followed by the literal binary sequence $s_i$ encoding the actual path. This description requires at most

$$l(s_i) + 2\log \delta_i \qquad (3)$$

bits. Concatenate the descriptions of all these paths into sequence $H$.

CLAIM 2 We can effectively reconstruct heap $h$ from $H$ and $n$.

PROOF. Assume $H$ is known and the fact that $h$ is a heap on $n$ different keys. We simulate the Sort step in reverse. Initially, $A[1..n]$ contains a sorted list with the least element in $A[1]$.

**for** $i := 2, \ldots, n-1$ **do:** {now $A[1..i-1]$ contains the partially constructed heap and $A[i..n]$ contains the remaining sorted list with the least element in $A[i]$} Put the key of $A[i]$ into $A[1]$, while shifting every key on the $(n-i)$th path in $H$ one position down starting from the root at $A[1]$. The last key on this path has nowhere to go and is put in the empty slot in $A[i]$.

**termination** {Array $A[1..n]$ contains heap $h$.}

$\square$

It follows from Claim 2 that $C(h|n) \leq l(H) + O(1)$. Therefore, by Equation 2, we have $l(H) \geq n \log n - 6n$. By the description in Equation 3, we have

$$\sum_{i=1}^{n}(l(s_i) + 2\log \delta_i) = \sum_{i=1}^{n}((\log n) - \delta_i + 2\log \delta_i) \geq n \log n - 6n.$$

It follows that $\sum_{i=1}^{n}(\delta_i - 2\log \delta_i) \leq 6n$. This is only possible if $\sum_{i=1}^{n} \delta_i = O(n)$. Therefore, the average path length is at least $\log n - c$, for some fixed constant $c$. In each round of the Sort step the path length equals the number of data moves. The combined total path length is at least $n \log n - nc$.

It follows that starting with heap $h$, Heapsort performs at least $n \log n - O(n)$ data moves. Trivially, the number of data moves is at most $n \log n$. Together this shows that Williams's method makes $2n \log n - O(n)$ key comparisons, and Floyd's method makes $n \log n + O(n)$ key comparisons.

Since most permutations are Kolmogorov random, these bounds for one random permutation $p$ also hold for all permutations *on average*. But we can make a stronger statement. We have taken $C(p|n)$ at least $\epsilon n$ below the possible maximum, for some constant $\epsilon > 0$. Hence, a fraction of at least $1 - 2^{-\epsilon n}$ of all permutations on $n$ keys will satisfy the above bounds. $\square$

## 4  Shellsort

The question of a nontrivial general lower bound (or upper bound) on the average complexity of Shellsort (due to D.L. Shell [26]) has been open for about four decades [11, 25], and only recently such a general lower bound was obtained. The original proof using Kolmogorov complexity [12] is presented here. Later, it turned out that the argument can be translated to a counting argument [13]. It is instructive that thinking in terms of code length and Kolmogorov complexity enabled advances in this problem.

Shellsort sorts a list of $n$ elements in $p$ passes using a sequence of increments $h_1, \ldots, h_p$. In the $k$th pass the main list is divided in $h_k$ separate sublists of length $\lceil n/h_k \rceil$, where the $i$th sublist consists of the elements at positions $j$, where $j \bmod h_k = i-1$, of the main list ($i = 1, \ldots, h_k$). Every sublist is sorted using a straightforward insertion sort. The efficiency of the method is governed by the number of passes $p$ and the selected increment sequence $h_1, \ldots, h_p$ with $h_p = 1$ to ensure sortedness of the final list. The original $\log n$-pass [3] increment sequence $\lfloor n/2 \rfloor, \lfloor n/4 \rfloor, \ldots, 1$ of Shell [26] uses worst case $\Theta(n^2)$ time, but Papernov and Stasevitch [19] showed that another related sequence uses $O(n^{3/2})$ and Pratt [22] extended this to a class of all nearly geometric increment sequences and proved this bound was tight. The currently best asymptotic method was found by Pratt [22]. It uses all $\log^2 n$ increments of the form $2^i 3^j < \lfloor n/2 \rfloor$ to obtain time $O(n \log^2 n)$ in the worst case. Moreover, since every pass takes at least $n$ steps, the average complexity using Pratt's increment sequence is $\Theta(n \log^2 n)$. Incerpi and Sedgewick [5] constructed a family of increment sequences for which Shellsort runs in $O(n^{1+\epsilon/\sqrt{\log n}})$ time using $(8/\epsilon^2) \log n$ passes, for every $\epsilon > 0$. B. Chazelle (attribution in [24]) obtained the same result by generalizing Pratt's method: instead of using 2 and 3 to construct the increment sequence use $a$ and $(a+1)$ to obtain a worst-case running time of $n \log^2 n (a^2/\ln^2 a)$ which is $O(n^{1+\epsilon/\sqrt{\log n}})$ for $\ln^2 a = O(\log n)$. Poonen [20], and Plaxton, Poonen and Suel [21], demonstrated an $\Omega(n^{1+\epsilon/\sqrt{p}})$ lower bound for $p$ passes of Shellsort using any increment sequence, for some $\epsilon > 0$; taking $p = \Omega(\log n)$ shows that the Incerpi-Sedgewick / Chazelle bounds are optimal for small $p$ and taking $p$ slightly larger shows a $\Theta(n \log^2 n / (\log \log n)^2)$ lower bound on the worst-case complexity of Shellsort. For the *average-case* running time Knuth [11] showed $\Theta(n^{5/3})$ for the best choice of increments in $p = 2$ passes; Yao [30] analyzed the average-case for $p = 3$ but did not obtain a simple analytic form; Yao's analysis was improved by Janson and Knuth [7] who showed $O(n^{23/15})$ average-case running time for a particular choice of increments in $p = 3$ passes. Apart from this no nontrivial results are known for the average-case; see [11, 24, 25]. In [12, 13] a general $\Omega(pn^{1+1/p})$ lower bound was obtained on the average-case running time of $p$-pass Shellsort under uniform distribution of input permutations, for every $1 \leq p \leq n/2$. [4] This is the first advance on the problem of determining general nontrivial bounds on the *average-case* running time of Shellsort [22, 11, 30, 5, 21, 24, 25].

A Shellsort computation consists of a sequence of comparison and inversion (swapping) operations. In this analysis of the average-case lower bound we count just the total number of data movements (here inversions) executed. The same bound holds *a fortiori* for the number of comparisons.

THEOREM 2 *The average number of comparisons (and also inversions for $p = o(\log n)$) in $p$-pass Shellsort on lists of $n$ keys is at least $\Omega\left(pn^{1+1/p}\right)$ for every increment sequence. The average is taken with all lists of $n$ items equally likely (uniform distribution).*

PROOF. Let the list to be sorted consist of a permutation $\pi$ of the elements $1, \ldots, n$. Consider a $(h_1, \ldots, h_p)$ Shellsort algorithm $A$ where $h_k$ is the increment in the $k$th pass and $h_p = 1$. For every $1 \leq i \leq n$ and $1 \leq k \leq p$, let $m_{i,k}$ be the number of elements in the $h_k$-increment sublist, containing element $i$, that are to the left of $i$ at the beginning of pass $k$ and are larger than $i$.

---

[3] "log" denotes the binary logarithm and "ln" denotes the natural logarithm.
[4] The trivial lower bound is $\Omega(pn)$ comparisons since every element needs to be compared at least once in every pass.

Observe that $\sum_{i=1}^{n} m_{i,k}$ is the number of inversions in the initial permutation of pass $k$, and that the insertion sort in pass $k$ requires precisely $\sum_{i=1}^{n}(m_{i,k}+1)$ comparisons. Let $M$ denote the total number of inversions:

$$M := \sum_{k=1}^{p} \sum_{i=1}^{n} m_{i,k}. \qquad (4)$$

CLAIM 3 *Given all the $m_{i,k}$'s in an appropriate fixed order, we can reconstruct the original permutation $\pi$.*

PROOF. In general, given the $m_{i,k}$'s and the final permutation of pass $k$, we can reconstruct the initial permutation of pass $k$. □

Let $M$ as in (4) be a fixed number. There are $n!$ permutations of $n$ elements. Let permutation $\pi$ be an incompressible permutation having Kolmogorov complexity

$$C(\pi|n, A, P) \geq \log n! - \log n, \qquad (5)$$

where $P$ is the decoding program in the following discussion. There exist many such permutations by lemma 1. Clearly, there is a fixed program that on input $A, P, n$ reconstructs $\pi$ from the description of the $m_{i,k}$'s as in Claim 3. Therefore, the minimum length of the latter description, including a fixed program in $O(1)$ bits, must exceed the complexity of $\pi$:

$$C(m_{1,1}, \ldots, m_{n,p}|n, A, P) + O(1) \geq C(\pi|n, A, P). \qquad (6)$$

An $M$ as defined by (4) such that every division of $M$ in $m_{i,k}$'s contradicts (6) would be a lower bound on the number of inversions performed. Similar to the reasoning Bubblesort example, Section 2, there are

$$D(M) := \binom{M + np - 1}{np - 1} \qquad (7)$$

distinct divisions of $M$ into $np$ ordered nonnegative integral summands $m_{i,k}$'s. Every division can be indicated by its index $j$ in an enumeration of these divisions. This is both obvious and an application of lemma 2. Therefore, a description of $M$ followed by a description of $j$ effectively describes the $m_{i,k}$'s. Fix $P$ as the program for the reference universal machine that reconstructs the ordered list of $m_{i,k}$'s from this description. The binary length of this two-part description must by definition exceed the Kolmogorov complexity of the described object.

A minor complication is that we cannot simply concatenate two binary description parts: the result is a binary string without delimiter to indicate where one substring ends and the other one begins. Encoding the $M$ part of the description self-delimitingly we obtain:

$$\log D(M) + \log M + 2 \log \log M + 1 \geq C(m_{1,1}, \ldots, m_{n,p}|n, A, P).$$

We know that $M \leq pn^2$ since every $m_{i,k} \leq n$. We can assume[5] $p < n$. Together with (5) and (6), we have

$$\log D(M) \geq \log n! - 4 \log n - 2 \log \log n - O(1). \qquad (8)$$

---
[5]Otherwise we require at least $n^2$ comparisons.

Estimate $\log D(M)$ by [6]

$$\log \binom{M+np-1}{np-1} = (np-1)\log\frac{M+np-1}{np-1} + M\log\frac{M+np-1}{M} + \frac{1}{2}\log\frac{M+np-1}{(np-1)M} + O(1).$$

The second term in the right-hand side equals[7]

$$\log\left(1+\frac{np-1}{M}\right)^M < \log e^{np-1}$$

for all positive $M$ and $np-1 > 0$. Since $0 < p < n$ and $n \leq M \leq pn^2$,

$$\frac{1}{2(np-1)}\log\frac{M+np-1}{(np-1)M} \to 0$$

for $n \to \infty$. Therefore, $\log D(M)$ is majorized asymptotically by

$$(np-1)\left(\log\left(\frac{M}{np-1}+1\right) + \log e\right)$$

for $n \to \infty$. Since the righthand-side of (8) is asymptotic to $n \log n$ for $n \to \infty$, this yields

$$M = \Omega(pn^{1+1/p}),$$

for $p = o(\log n)$. (More precisely, $M = \Omega(pn^{1+(1-\epsilon)/p})$ for $p \leq (\epsilon/\log e)\log n$ ($0 < \epsilon < 1$), see [13].) That is, the running time of the algorithm is as stated in the theorem for every permutation $\pi$ satisfying (5). By lemma 1 at least a $(1-1/n)$-fraction of all permutations $\pi$ require that high complexity. Then the following is a lower bound on the expected number of inversions of the sorting procedure:

$$(1-\frac{1}{n})\Omega(pn^{1+1/p}) + \frac{1}{n}\Omega(0) = \Omega(pn^{1+1/p}),$$

for $p = o(\log n)$. For $p = \Omega(\log n)$, the lower bound on the number of comparisons is trivially $pn = \Omega(pn^{1+1/p})$. This gives us the theorem. □

Our lower bound on the average-case can be compared with the Plaxton-Poonen-Suel $\Omega(n^{1+\epsilon/\sqrt{p}})$ worst case lower bound [21]. Some special cases of the lower bound on the average-case complexity are:

1. For $p = 1$ our lower bound is asymptotically tight: it is the average number of inversions for Insertion Sort.

2. For $p = 2$, Shellsort requires $\Omega(n^{3/2})$ inversions (the tight bound is known to be $\Theta(n^{5/3})$ [11]);

3. For $p = 3$, Shellsort requires $\Omega(n^{4/3})$ inversions (the best known upper bound is $O(n^{23/15})$ in [7]);

---

[6]Use the following formula ([16], p. 10),

$$\log\binom{a}{b} = b\log\frac{a}{b} + (a-b)\log\frac{a}{a-b} + \frac{1}{2}\log\frac{a}{b(a-b)} + O(1).$$

[7]Use $e^a > (1+\frac{a}{b})^b$ for all $a > 0$ and positive integer $b$.

4. For $p = \log n / \log \log n$, Shellsort requires $\Omega(n \log^2 n / \log \log n)$ inversions;

5. For $p = \log n$, Shellsort requires $\Omega(n \log n)$ comparisons on average. This is of course the lower bound of average number of comparisons for every sorting algorithm.

6. In general, for $n/2 > p = p(n) > \log n$, Shellsort requires $\Omega(n \cdot p(n))$ comparisons (since every pass trivially makes $n$ comparisons).

In [25] it is mentioned that the existence of an increment sequence yielding an average $O(n \log n)$ Shellsort has been open for 30 years. The above lower bound on the average shows that the number $p$ of passes of such an increment sequence (if it exists) is precisely $p = \Theta(\log n)$; all the other possibilities are ruled out: Is there an increment sequence for $\log n$-pass Shellsort so that it runs in average-case $\Theta(n \log n)$? Can we tighten the average-case lower bound for Shellsort? The above bound is known to be not tight for $p = 2$ passes.

## 5 Dobosiewicz Sort and Shakersort

We look at some variants of Shellsort. Knuth [11], 1st Edition Exercise 5.2.1.40 on page 105, and Dobosiewicz [3] proposed to use only one pass of Bubblesort on each subsequence instead of sorting the subsequences at each stage. Incerpi and Sedgewick [6] used two passes of Bubblesort in each stage, one going left-to-right and the other going right-to-left. This is called Shakersort since it reminds one of shaking a cocktail. In both cases the sequence may stay unsorted, even if the last increment is 1. A final phase, a straight insertion sort, is required to guaranty a fully sorted list. Until recently, these variants have not been seriously analyzed; in [3, 6, 28, 24] mainly empirical evidence is reported, giving evidence of good running times (comparable to Shellsort) on randomly generated input key sequences of moderate length. The evidence also suggests that the worst-case running time may be quadratic. Again, let $n$ be the number of keys to be sorted and let $p$ be the number of passes. The $\Omega(n^{1+c/\sqrt{p}})$ worst-case lower bound of Poonen [20] holds apart from Shellsort also for the variants of it. We also have a worst-case lower bound of $\Omega(n^2)$ on Dobosiewicz sort and Shaker sort for the special case of almost geometric sequences of increments. But recently Brejova [1] proved that Shaker sort runs in $O(n^{3/2} \log^3 n)$ worst-case time for a certain sequence of increments (the first non-quadratic worst-case upper bound). Using the incompressibility method, she also proved lower bounds on the average-case running times.

THEOREM 3 *There is an $\Omega(n^2/4^p)$ lower bound on the average-case running time of Shaker sort, and a $\Omega(n^2/2^p)$ lower bound on the average-case running time of Dobosiewicz sort. The avereges are taken with respect to the uniform distribution.*

REMARK 1 These lower bounds (on the average-case) are better than the Poonen [20] lower bounds of $\Omega(n^{1+c/\sqrt{p}})$ on the worst-case.

PROOF. Consider Dobosiewicz sorting algorithm $A$ (the description of $A$ includes the number of passes $p$ and the list of increments $h_1, \ldots, h_p$). Every comparison based sorting algorithm uses $\Omega(n \log n)$ comparisons on average. If $p > \log n - \log \log n$ then the claimed lower bound trivially holds. So we can assume that $p \leq \log n - \log \log n$. Let $\pi$ be the permutation of $\{0, 1, \ldots, n-1\}$ to be sorted, and let $\pi'$ be the permutation remaining after all $p$ stages of the Dobsiewicz sort, but before the final insertion sort. If $X$ is the number of inversions in $\pi'$ then the final insertion sort takes $\Omega(X)$ time.

CLAIM 4 *Let $\pi$ be a permutation satisfying (5). Then $X = \Omega(n^2/2^p)$.*

PROOF. We can reconstruct $\pi$ from $\pi'$ given $p$ strings of lengths $n$ defined as follows: The $j$th bit of the $i$th string is "1" if $x_j$ was interchanged with $x_{j-h_i}$ in the $i$th phase of the algorithm ($h_i$ is the $i$th increment), and "0" otherwise. Given $\pi'$ and these strings in appropriate order we can simply run the $p$ sorting phases in reverse.

Furthermore, $\pi'$ can be reconstructed from its inversion table $a_0, a_2, \ldots, a_{n-1}$, where $a_i$ is the number of elements in list $\pi'$ left of the $i$th position that are greater than the element in the $i$th position. Thus, $\sum_i a_i = X$. There are $D(X) = \binom{X+n-1}{n-1}$ ordered partitions of $X$ into $n$ non-negative summands. Hence, $\pi'$ can be reconstructed from $X$ and an index of $\log D(X)$ bits identifying the partition in question. Given $n$, we encode $X$ self-delimiting to obtain a total description of $\log X + 2 \log \log X + \log D(X)$ bits.

Therefore, with $P$ the reconstruction program, we have shown that

$$C(\pi \mid n, A, P) \leq np + \log X + 2 \log \log X + \log D(X).$$

Estimating asymptotically, similar to the part following (8),

$$\log D(X) \leq (n-1) \log \left( \frac{X}{n-1} + 1 \right) + O(n).$$

Since $\pi$ satisfies (5), we have $np + (n-1) \log \left( \frac{X}{n-1} + 1 \right) + O(n) \geq n \log n - \Theta(n)$. Hence, $X \geq n^2/(2^p)\Theta(1) = \Omega(n^2/2^p)$, where the last equality holds since $p \leq \log n - \log \log n$ and hence $n^2/2^p \geq n \log n$. □

By lemma 1 at least a $(1 - 1/n)$-fraction of all permutations $\pi$ require that high complexity. This shows that the running time of the Dobosiewicz sort is as stated in the theorem. The lower bound on Shaker sort has a very similar proof, with the proviso that we require $2n$ bits to encode one pass of the algorithm rather than $n$ bits. This results in the claimed lower bound of $\Omega(n^2/4^p)$ (which is nan-vacuous only for for $p \leq \frac{1}{2}(\log n - \log \log n)$). □

## 6 Sorting with Queues and Stacks

Knuth [11] and Tarjan [27] have studied the problem of sorting using a network of queues or stacks. The main variant of the problem is as follws: Given that the stacks or queues are arranged sequentially as shown in Figure 1, or in parallel as shown in Figure 2. Question: how many stacks or queues are needed to sort $n$ elements with comparisons only? We assume that the input sequence is scanned from left to right, and the elements follow the arrows to go to the next stack or queue or output. In [12, 14] only the average-case analyses of the above two main variants was given, although the technique applies more in general to arbitrary acyclic networks of stacks and queues as studied in [27].

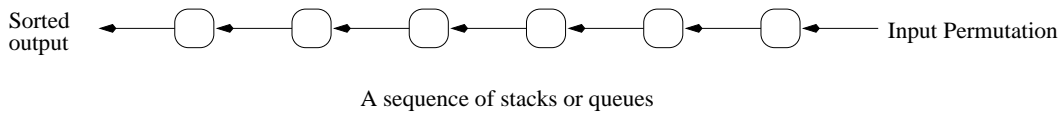

Figure 1: Six stacks/queues arranged in sequential order

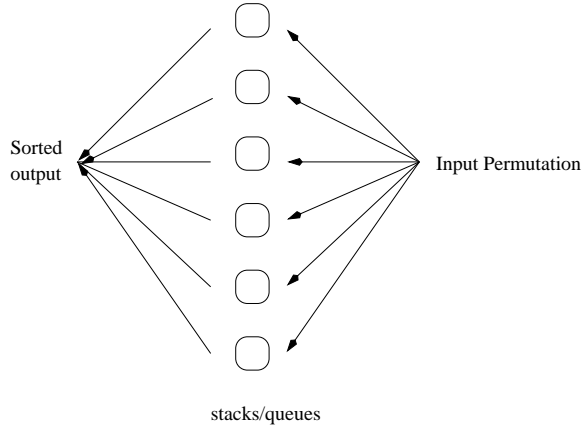

Figure 2: Six stacks/queues arranged in parallel order

## 6.1 Sorting with Sequential Stacks

The sequential stack sorting problem is given in [11] exercise 5.2.4-20. We have $k$ stacks numbered $S_0, \ldots, S_{k-1}$. The input is a permutation $\pi$ of the elements $1, \ldots, n$. Initially we push the elements of $\pi$ on $S_0$, at most one at a time, in the order in which they appear in $\pi$. At every step we can pop a stack (the popped elements will move left in Figure 1) or push an incoming element on a stack. The question is how many stack are needed for sorting $\pi$. It is known that $k = \log n$ stacks suffice, and $\frac{1}{2} \log n$ stacks are necessary in the worst-case [11, 27]. Here we prove that the same lower bound also holds on the average, using a very simple incompressibility argument.

THEOREM 4 *On average (uniform distribution), at least $\frac{1}{2} \log n$ stacks are needed for sequential stack sort.*

PROOF. Fix a random permutation $\pi$ such that

$$C(\pi|n, P) \geq \log n! - \log n = n \log n - O(n),$$

where $P$ is an encoding program to be specified in the following.

Assume that $k$ stacks are sufficient to sort $\pi$. We now encode such a sorting process. For every stack, exactly $n$ elements pass through it. Hence we need perform precisely $n$ pushes and $n$ pops on every stack. Encode a push as 0 and a pop as 1. It is easy to prove that different permutations must have different push/pop sequences on at least one stack. Thus with $2kn$ bits, we can completely specify the input permutation $\pi$. Then, as before,

$$2kn \geq \log n! - \log n = n \log n - O(n).$$

Therefore, we have $k \geq \frac{1}{2} \log n - O(1)$ for the random permutation $\pi$.

Since most (a $(1 - 1/n)$th fraction) permutations are incompressible, we calculate the average-case lower bound as:

$$\frac{1}{2} \log n \cdot \frac{n-1}{n} + 1 \cdot \frac{1}{n} \approx \frac{1}{2} \log n.$$

□

## 6.2 Sorting with Parallel Stacks

Clearly, the input sequence $2, 3, 4, \ldots, n, 1$ requires $n - 1$ parallel stacks to sort. Hence the worst-case complexity of sorting with parallel stacks, as shown in Figure 2, is $n - 1$. However, most sequences do not need this many stacks to sort in a parallel arrangement. The next two theorems show that on average, $\Theta(\sqrt{n})$ stacks are both necessary and sufficient. Observe that the result is actually implied by the connection between sorting with parallel stacks and *longest increasing subsequences* in [27], and the bounds on the length of longest increasing subsequences of random permutations given in [9, 17, 8]. However, the proofs in [9, 17, 8] use deep results from probability theory (such as Kingman's ergodic theorem) and are quite sophisticated. Here we give simple proofs using incompressibility arguments.

THEOREM 5 *On average (uniform distribution), the number of parallel stacks needed to sort $n$ elements is $O(\sqrt{n})$.*

PROOF. Consider an incompressible permutation $\pi$ satisfying

$$C(\pi|n) \geq \log n! - \log n. \tag{9}$$

We use the following trivial algorithm (described in [27]), to sort $\pi$ with stacks in the parallel arrangement shown in Figure 2. Assume that the stacks are $S_0, S_1, \ldots$, and the input sequence is denoted as $x_1, \ldots, x_n$.

**Algorithm Parallel-Stack-Sort**

1. For $i = 1$ to $n$ do

   Scan the stacks from left to right, and push $x_i$ on the the first stack $S_j$ whose top element is larger than $x_i$. If such a stack doesn't exist, put $x_i$ on the first empty stack.

2. Pop the stacks in the ascending order of their top elements.

We claim that algorithm Parallel-Stack-Sort uses $O(\sqrt{n})$ stacks on the permutation $\pi$. First, we observe that if the algorithm uses $m$ stacks on $\pi$ then we can identify an increasing subsequence of $\pi$ of length $m$ as in [27]. This can be done by a trivial backtracking starting from the top element of the last stack. Then we argue that $\pi$ cannot have an increasing subsequence of length longer than $e\sqrt{n}$, where $e$ is the natural constant, since it is compressible by at most $\log n$ bits.

Suppose that $\sigma$ is a longest increasing subsequence of $\pi$ and $m = |\sigma|$ is the length of $\sigma$. Then we can encode $\pi$ by specifying:

1. a description of this encoding scheme in $O(1)$ bits;

2. the number $m$ in $\log m$ bits;

3. the combination $\sigma$ in $\log \binom{n}{m}$ bits;

4. the locations of the elements of $\sigma$ in $\pi$ in at most $\log \binom{n}{m}$ bits; and

5. the remaining $\pi$ with the elements of $\sigma$ deleted in $\log(n - m)!$ bits.

This takes a total of

$$\log(n-m)! + 2\log\frac{n!}{m!(n-m)!} + \log m + O(1) + 2\log\log m$$

bits, where the last $\log\log m$ term serves to self-delimitingly encode $m$. Using Stirling's approximation, and the fact that $\sqrt{n} \leq m = o(n)$, the above expression is upper bounded by:

$$\log n! + \log\frac{(n/e)^n}{(m/e)^{2m}((n-m)/e)^{n-m}} + O(\log n)$$
$$\approx \log n! + m\log\frac{n}{m^2} + (n-m)\log\frac{n}{n-m} + m\log e + O(\log n)$$
$$\approx \log n! + m\log\frac{n}{m^2} + 2m\log e + O(\log n)$$

This description length must exceed the complexity of the permutation which is lower-bounded in (9). Therefore, approximately $m \leq e\sqrt{n}$, and hence $m = O(\sqrt{n})$. Hence, the average complexity of Parallel-Stack-Sort is

$$O(\sqrt{n}) \cdot \frac{n-1}{n} + n \cdot \frac{1}{n} = O(\sqrt{n}).$$

□

THEOREM 6 *On average (uniform distribution), the number of parallel stacks required to sort a permutation is $\Omega(\sqrt{n})$.*

PROOF. Let $A$ be a sorting algorithm using parallel stacks. Fix a random permutation $\pi$ with $C(\pi|n, P) \geq \log n! - \log n$, where $P$ is the program to do the encoding discussed in the following. Suppose that $A$ uses $T$ parallel stacks to sort $\pi$. This sorting process involves a sequence of moves, and we can encode this sequence of moves by a sequence of instructions of the types:

- push to stack $i$,
- pop stack $j$,

where the element to be pushed is the next unprocessed element from the input sequence, and the popped element is written as the next output element. Each of these term requires $\log T$ bits. In total, we use precisely $2n$ terms since every element has to be pushed once and has to be popped once. Such a sequence is unique for every permutation.

Thus we have a description of an input sequence in $2n\log T$ bits, which must exceed $C(\pi|n, P) \geq n\log n - O(\log n)$. It follows that $T \geq \sqrt{n} = \Omega(\sqrt{n})$.

This yields the average-case complexity of $A$:

$$\Omega(\sqrt{n}) \cdot \frac{n-1}{n} + 1 \cdot \frac{1}{n} = \Omega(\sqrt{n}).$$

□

## 6.3 Sorting with Parallel Queues

It is easy to see that sorting cannot be done with a sequence of queues. So we consider the complexity of sorting with parallel queues. It turns out that all the result in the previous subsection also hold for queues.

As noticed in [27], the worst-case complexity of sorting with parallel queues is $n$, since the input sequence $n, n-1, \ldots, 1$ requires $n$ queues to sort. We show in the next two theorems that on average, $\Theta(\sqrt{n})$ queues are both necessary and sufficient. Again, the result is implied by the connection between sorting with parallel queues and longest *decreasing* subsequences given in [27] and the bounds in [9, 17, 8] (with sophisticated proofs). Our proofs are trivial given the proofs in the previous subsection.

THEOREM 7 *On average (uniform distribution), the number of parallel queues needed to sort $n$ elements is upper bounded by $O(\sqrt{n})$.*

PROOF. The proof is very similar to the proof of Theorem 5. We use a slightly modified greedy algorithm as described in [27]:

**Algorithm Parallel-Queue-Sort**

1. For $i = 1$ to $n$ do

    Scan the queues from left to right, and append $x_i$ on the the first queue whose rear element is smaller than $x_i$. If such a queue doesn't exist, put $x_i$ on the first empty queue.

2. Delete the front elements of the queues in the ascending order.

Again, we claim that algorithm Parallel-Queue-Sort uses $O(\sqrt{n})$ queues on every permutation $\pi$, that cannot be compressed by more than $\log n$ bits. We first observe that if the algorithm uses $m$ queues on $\pi$ then a decreasing subsequence of $\pi$ of length $m$ can be identified, and we then argue that $\pi$ cannot have a decreasing subsequence of length longer than $e\sqrt{n}$, in a way analogous to the argument in the proof of Theorem 5.

□

THEOREM 8 *On average (uniform distribution), the number of parallel queues required to sort a permutation is $\Omega(\sqrt{n})$.*

PROOF. The proof is the same as the one for Theorem 6 except that we should replace "push" with "enqueue" and "pop" with "dequeue". □

# References

[1] B. Brejová, Analyzing variants of Shellsort, *Information Processing Letters*, 79:5(2001), 223–227.

[2] H. Buhrman, T. Jiang, M. Li, and P. Vitányi, New applications of the incompressibility method, pp. 220–229 in *the Proceedings of ICALP'99*, LNCS 1644, Springer-Verlag, Berlin, 1999.

[3] W. Dobosiewicz, An efficient variant of bubble sort, *Information Processing Letters*, 11:1(1980), 5–6.

[4] R.W. Floyd, Algorithm 245: Treesort 3. *Communications of the ACM*, 7(1964), 701.

[5] J. Incerpi and R. Sedgewick, Improved upper bounds on Shellsort, *Journal of Computer and System Sciences*, 31(1985), 210–224.

[6] J. Incerpi and R. Sedgewick, Practical variations of Shellsort, *Information Processing Letters*, 26:1(1980), 37–43.

[7] S. Janson and D.E. Knuth, Shellsort with three increments, *Random Struct. Alg.*, 10(1997), 125-142.

[8] S.V. Kerov and A.M. Versik, Asymptotics of the Plancherel measure on symmetric group and the limiting form of the Young tableaux, *Soviet Math. Dokl.* 18 (1977), 527-531.

[9] J.F.C. Kingman, The ergodic theory of subadditive stochastic processes, *Ann. Probab.* 1 (1973), 883-909.

[10] A.N. Kolmogorov, Three approaches to the quantitative definition of information. *Problems Inform. Transmission*, 1:1(1965), 1–7.

[11] D.E. Knuth, *The Art of Computer Programming, Vol.3: Sorting and Searching*, Addison-Wesley, 1973 (1st Edition), 1998 (2nd Edition).

[12] T. Jiang, M. Li, and P. Vitanyi, Average complexity of Shellsort (preliminary version), *Proc. ICALP99*, Lecture Notes in Computer Science, Vol. 1644, Springer-Verlag, Berlin, 1999, 453–462.

[13] T. Jiang, M. Li, and P. Vitanyi, A lower bound on the average-case complexity of Shellsort, *J. Assoc. Comp. Mach.*, 47:5(2000), 905–911.

[14] T. Jiang, M. Li, and P. Vitanyi, Average-case analysis of algorithms using Kolmogorov complexity, *Journal of Computer Science and Technology*, 15:5(2000), 402–408.

[15] T. Jiang, M. Li, and P. Vitányi, The average-case area of Heilbronn-type triangles, *Random Structures and Algorithms*, 20:2(2002), 206-219.

[16] M. Li and P.M.B. Vitányi, *An Introduction to Kolmogorov Complexity and its Applications*, Springer-Verlag, New York, 2nd Edition, 1997.

[17] B.F. Logan and L.A. Shepp, A variational problem for random Young tableaux, *Advances in Math.* 26 (1977), 206-222.

[18] I. Munro, Personal communication, 1992.

[19] A. Papernov and G. Stasevich, A method for information sorting in computer memories, *Problems Inform. Transmission*, 1:3(1965), 63–75.

[20] B. Poonen, The worst-case of Shellsort and related algorithms, *J. Algorithms*, 15:1(1993), 101-124.

[21] C.G. Plaxton, B. Poonen and T. Suel, Improved lower bounds for Shellsort, *Proc. 33rd IEEE Symp. Foundat. Comput. Sci.*, pp. 226–235, 1992.

[22] V.R. Pratt, *Shellsort and Sorting Networks*, Ph.D. Thesis, Stanford Univ., 1972.

[23] R. Schaffer and R. Sedgewick, *J. Algorithms*, 15(1993), 76–100.

[24] R. Sedgewick, Analysis of Shellsort and related algorithms, presented at the *Fourth Annual European Symposium on Algorithms*, Barcelona, September, 1996.

[25] R. Sedgewick, Open problems in the analysis of sorting and searching algorithms, Presented at *Workshop on Prob. Analysis of Algorithms*, Princeton, 1997.

[26] D.L. Shell, A high-speed sorting procedure, *Commun. ACM*, 2:7(1959), 30–32.

[27] R.E. Tarjan, Sorting using networks of queues and stacks, *Journal of the ACM*, 19(1972), 341–346.

[28] M.A. Weiss and R. Sedgewick, Bad cases for Shaker-sort, *Information Processing Letters*, 28:3(1988), 133–136.

[29] J.W.J. Williams *Comm. ACM*, 7(1964), 347–348.

[30] A.C.C. Yao, An analysis of $(h, k, 1)$-Shellsort, *J. of Algorithms*, 1(1980), 14–50.